\documentclass[sts]{imsart}

\usepackage{algorithm,algorithmic,amsfonts,amssymb,amsmath,amscd,amsthm,bm,latexsym}
\usepackage{natbib}
\usepackage{graphicx}
\usepackage{picins}
\usepackage{upgreek}
\usepackage{units}
\usepackage{nicefrac}

\renewcommand{\rho}{\varrho}

\begin{document}

\begin{frontmatter}

\title{Some comments about A.~Ronald Gallant's ``Reflections on the probability space induced by moment conditions with
implications for {B}ayesian inference"}
\runtitle{Discussion of R. Gallant's paper}
\thankstext{T1}{Christian P. Robert, CEREMADE, Universit{\' e} Paris-Dauphine, 75775 Paris cedex 16, France
{\sf xian@ceremade.dauphine.fr}. Research partly supported by the Agence Nationale de la Recherche (ANR,
212, rue de Bercy 75012 Paris) through the 2012--2015 grant ANR-11-BS01-0010 ``Calibration'' and by a Institut 
Universitaire de France senior chair. C.P.~Robert is also affiliated as a part-time professor in the Department of
Statistics of the University of Warwick, Coventry, UK.}

\begin{aug}
\author{\snm{Christian P.~Robert}}
\affiliation{Universit{\'e} Paris-Dauphine, CEREMADE, and CREST, Paris}
\end{aug}

\begin{abstract}
This note is commenting on Ronald Gallant's (2015) reflections on the construction of Bayesian prior distributions from moment
conditions. The main conclusion is that the paper does not deliver a working principle that could justify inference
based on such priors.
\end{abstract}

\end{frontmatter}

\section{Introduction}

The construction of prior distributions has always been a central aspect of Bayesian analysis, arguably {\em the}
central piece since all aspects of Bayesian inference are automatically derived from defining the model, the prior, the
data and the loss function \citep{berger:1985}. While the prior is a mathematical object, there is not rigorous
derivation of a given prior distribution from the available information or lack thereof and, while ``objective Bayes"
constructions are automated to some extent \citep{lhoste:1923,jeffreys:1939,broemeling:2003,berger:bernardo:sun:2009},
they are rejected by subjectivist Bayesians who argue in favour of personalistic and non-reproducible prior selection
\citep{kadane:2011}.  Defining priors via moment conditions can be traced at least back to \cite{jaynes:2003} and the
notion of {\em maximum entropy priors}, even though the moment conditions only involve the parameters $\theta$ of the
model. The current paper considers instead moment conditions as defined jointly on the pair $(x,\theta)$ and proposes
some necessary conditions for a prior distribution to be compatible with those conditions. From a foundational
perspective, a setting where the joint distribution of the data and the parameter that drives this data
is a given is hard to fathom, because it implies there is no longer a fixed parameter to infer about.
In addition, the term ``exogeneity" used in the paper hints at a notion of the parameter being not truly a parameter,
but including rather latent variables and maybe random effects. It is hard to reconcile this motivation with a
computational one (also found in the paper) where complex likelihoods would justify calling for a prior as a practical
tool. The additional reformulation of a (pseudo-)likelihood function defined by the method of moments (Section 2) makes the 
focus of the paper difficult to specify and hence analyse. 

Given the introduction through Fisher's (1930) fiducial distribution, one may wonder whether or not the author's
approach is to integrate fiducial constructs within the Bayesian paradigm or at the very least to specify under which
conditions this can be achieved. As a long-time sceptic on the relevance of fiducial arguments, I doubt
about one's ability to produce an arbitrary distribution on an equally arbitrary transform of the pair $(x,\theta)$,
instead of a genuine prior $\times$ likelihood construct. For instance, the discussion around the various meanings of
the $t$ statistic
$$\dfrac{\bar{x}-\theta}{\nicefrac{s}{\sqrt{n}}}$$
does not imply that it can achieve a $t$ posterior distribution with $n-1$ degrees of freedom {\em jointly} for all
sample sizes $n$ and I doubt it can happen outside exotic cases like Dirac masses on one of the terms. (Exchanging the
randomness of terms in a random variable as if it were a linear equation is a guaranteed way to produce fiducial
paradoxes and measure theoretic difficulties.)

This set of comments on \cite{gallant:2015} is organised as follows: in Section \ref{sec:moma}, I analyse various and
somewhat mutually exclusive aspects of the author's approach, while in Section \ref{sec:compadre}, I discuss some computational
consequences and alternatives.

\section{Deriving priors from moments}\label{sec:moma}

\subsection{Moment default likelihood}

Gallant (2015) considers the distribution of a pivotal quantity like
$$
Z=\sqrt{n} W(\mathbf{x},\theta)^{-\nicefrac{1}{2}} m(\mathbf{x},\theta)
$$
as induced by the hypothetical joint distribution on $(x,\theta)$, hence conversely inducing constraints on this joint,
as well as an associated conditional. (The constraints may be such that the joint distribution does not exist.) However,
this perspective is abandoned a few lines below to define a moment likelihood
$$
p(x | \theta) = (2\pi)^{-\nicefrac{M}{2}} \exp \left\{ \nicefrac{-n}{2}\,
\bar{m}(x, \theta)^\text{T}[W(x, \theta)]^{-1} \bar{m}(x, \theta) \right\}
$$
as a quasi-Gaussian pseudo-likelihood in the moment $\bar{m}(x, \theta)$. This is only one among many ways of defining a
likelihood from moments, but it further removes the symmetry in $x$ and $\theta$ induced by the original formulation. In
addition, one may wonder why a determinant like $\text{det}\{W(x, \theta)\}^{\nicefrac{-1}{2}}$ or at least a normalising
constant (obviously depending on $\theta$) does not appear in the pseudo-likelihood, since this impacts the resulting posterior
density.

A connected reference is Zellner's (1997) {\em Bayesian method of moments} where, given moment conditions on the
parameters $\theta$ and $\sigma^2$,
$$
\mathbb{E}[\theta|x_1,\ldots,x_n] = \bar{x}_n\,,
\quad
\mathbb{E}[\sigma^2|x_1,\ldots] = s^2_n\,,
\quad
\text{var}(\theta|\sigma^2,x_1,\ldots) = \nicefrac{\sigma^2}{n}\,,
$$
\cite{zellner:1997} derives a {\em maximum entropy} posterior
$$
\theta|\sigma^2,x_1,\ldots\sim\mathcal{N}(\bar{x}_n,\nicefrac{\sigma^2}{n})\,,
\quad
\sigma^{-2}|x_1,\ldots\sim\mathcal{E}xp(s^2_n)\,,
$$
later shown to be incompatible with the corresponding predictive distribution, besides producing an inconsistent
estimator of $\sigma^2$ \citep{geisser:1999}.\footnote{In essence, the prior changes for each sample size $n$.
\cite{geisser:1999} designates the Maxent principle {\em per se} as the culprit for this incoherence.}

\subsection{Measure-theoretic considerations}

    \begin{quote}``If one specifies a set of moment functions collected together into a vector $m(x,\theta)$ of
dimension $M$, regards $\theta$ as random and asserts that some transformation $Z(x,\theta)$ has distribution $\psi$
then what is required to use this information and then possibly a prior to make valid inference?" (p.4)\end{quote}

The central question in the paper is determining whether a set of moment equations
\begin{equation}\label{eq:mom}
\mathbb{E}[m(X_1,\ldots,X_n,\theta)]=0
\end{equation}
(where both the $X_i$‘s and $\theta$ are {\em a priori} random) leads to a well-defined pair of a likelihood function and a prior
distribution compatible with those. From a mathematical perspective, this seems to be a highly complex question 
as it implies the integral equation
$$
\int_{\Theta\times\mathcal{X}^n} m(x_1,\ldots,x_n,\theta)\,\pi(\theta)f(x_1|\theta)\cdots f(x_n|\theta)
\text{d}\theta\,\text{d}x_1\cdots\text{d}x_n=0
$$
must allow for a solution {\em for all} $n$'s.

Still from a purely mathematical perspective, the problem as stated in Section 3.3 of Gallant (2015) is puzzling: 
if the distribution of the transform $Z=Z(X,\Lambda)$ is provided, what are the consequences on the joint distribution
of $(X,\Lambda)$? It is conceivable but rather unlikely that this distribution $\psi$ will induce a single joint, that
is, a single prior and a single likelihood. It is much more likely that the distribution $\psi$ one arbitrarily selects on 
$m(x,\theta)$ is incompatible with a joint distribution on $(x,\theta)$. To wit, Fisher's example of the $t$ statistic
and of its $t_{n-1}$ distribution.

    \begin{quote}``Typically $C$ is coarse in the sense that it does not contain all the Borel sets (...)  
The probability space cannot be used for Bayesian inference." (p.8) \end{quote}

My understanding of that part of the paper is that defining a joint on $m(x,\theta)$ is not always enough to deduce a
(unique) posterior on $\theta$, which is fine and correct, but definitely anticlimactic. This sounds to be what Gallant
calls a ``partial specification of the prior" (p.9). Hence, rather than building the minimal Borel $\sigma$-algebra on
$\mathcal{X}\times\Theta$ compatible with this joint on $m(x,\theta)$, I would suggest examining the range of
prior$\times$likelihood pairs that agree with this partial property when using the regular Borel $\sigma$-algebra.

The general solution found in Section 3.5 (``The Abstraction") relies on the assumptions that $Z(\cdot,\theta)$ is
a surjective function for all $\theta$'s and on the axiom of choice, namely that an antecedent of the function can be
selected for each $z\in\mathcal{Z}$, namely, $\Upsilon(z,\theta)=x^\star$ such that $Z(x^\star,\theta)=z$. Under these
assumptions, $Z$ and $\Upsilon(Z,\theta)$ are in one-to-one correspondence and hence can enjoy the same distribution
modulo the proper change of variable. The distribution over $X$ is then obtained by assuming a uniform distribution over
the orbit of $Z(x,\theta)=Z(x^*,\theta)$, leading to 
$$p^\star(x|\theta) = \psi(Z(x,\theta))$$
as defined in Equation (17). There is no issue about this derivation but, as noted previously, there is neither a
compelling reason to adopt the smallest $\sigma$-algebra $\mathcal{C}^*$ to make the above a proper density in
$\mathcal{X}$. I see little appeal in using this new measure and further wonder in which sense this defines 
a likelihood function, i.e., the product of $n$ densities of the $X_i$‘s conditional on $\theta$. To me this is the
central issue, which remains unsolved by the paper.

\subsection{Computational motivations}

    \begin{quote}``A common situation that requires consideration of the notions that follow is that deriving the
likelihood from a structural model is analytically intractable and one cannot verify that the numerical approximations
one would have to make to circumvent the intractability are sufficiently accurate." (p.7)\end{quote}

This computational perspective then is a completely different issue, namely that defining a joint distribution by
mean of moment equations prevents regular Bayesian inference when the likelihood function is intractable. This point of
view is much more exciting because (i) there are alternatives available, from approximate Bayesian computation (ABC)
\citep{marin:pudlo:robert:ryder:2011} to INLA \citep{rue:martino:chopin:2008}, to EP \citep{barthelme:chopin:2014}, to
variational Bayes \citep{jaakkola:jordan:2000}.  In particular, the moment equations are strongly and even insistently
suggesting that empirical likelihood techniques \citep{owen:2001,lazar:2003} could be well-suited to this setting. And
(ii) it is no longer a mathematical puzzle: there exists a joint distribution on $m(x,\theta)$, induced by one (or many)
joint distribution(s) on $(x,\theta)$. Hence, the question of finding whether or not this item of information leads to a
single proper prior on $\theta$ becomes irrelevant. However, in the event one wants to rely on ABC, being given the
distribution of $m(x,\theta)$ seems to mean one can solely generate new values of this transform while missing a natural
distance between observations and pseudo-observations, although the log-likelihood of $m(x^\text{obs},\theta)$ could
itself be used as a distance. 

As an aside, the author mentions marginal likelihood estimation by harmonic means \`a la \cite{newton:raftery:1994}, but
I would like to point out this usually is a rather poor solution with potential for disaster, while it requires the
likelihood function to be available in closed form. It is also unclear to me why marginal likelihood is mentioned at
this stage.

\section{A form of ABC?}\label{sec:compadre}

    \begin{quote}``“These characteristics are (1) likelihood is not available; (2) prior information is available; (3) a
portion of the prior information is expressed in terms of functionals of the model that cannot be converted into an
analytic prior on model parameters; (4) the model can be simulated. Our approach depends on an assumption that (5) an
adequate statistical model for the data are available." R. Gallant and R.  McCulloch (2009)\end{quote}

As a final comment connected with the computational aspect of the current paper, I would like to point out Gallant's and McCulloch's (2009) connections with the ABC approach, to wit the above quote.

In \cite{gallant:mcculloch:2009}, the true (scientific) model parametrised by $\theta$ is replaced with a (statistical)
substitute that is available in closed form and parametrised by $g(\theta)$.  which states that the intractable density
is equal to a closed-form density.] This latter model is over-parametrised when compared with the scientific model.
Take, e.g., a $\mathcal{N}(\theta,\theta^2)$ scientific model versus a $\mathcal{N}(\mu,\sigma^2)$ statistical model. In
addition, the prior information is only available on the parameter $\theta$. However, this does not seem to matter very
much since (a) the Bayesian analysis is operated on $\theta$ only and (b) the Metropolis approach adopted by the authors
involves simulating a massive number of pseudo-observations, given the current value of the parameter $\theta$ and the
scientific model, so that the transform $g(\theta)$ can be estimated by maximum likelihood over the statistical model.
The paper suggests using a secondary Markov chain algorithm to find this MLE. The pseudo-model is then used in a primary MCMC step.

Hence, the approach of \cite{gallant:mcculloch:2009} is not truly an ABC algorithm. In the same setting, ABC would
indeed use one simulated dataset, with the same size as the observed dataset, compute the MLEs for both and compare them
(as in \citealp{drovandi:pettitt:faddy:2011,martin:etal:2010}). This approach is faster if less accurate when Assumption
1---that the statistical model holds for a restricted parametrisation---does not stand. 

\section{Conclusion}

One overall interrogation about this paper is the validation of the outcome. As noted in \cite{fraser:2011}, Bayesian
posterior distributions are not naturally endowed with an epistemic validity. The same questioning obviously applies to
entities defined outside the Bayesian paradigm, the present one included, in that producing a posterior or
pseudo-posterior distribution on the parameter offers no guarantee {\em per se} about the efficiency of the inference 
it produces. Using asymptotically convergent approximations to the likelihood function does not always lead to
consistent Bayesian approximations \citep{marin:pillai:robert:rousseau:2011} and thus requires further validation of the
procedures proposed here.

Another global interrogation that remains open is the validation of the income outside of the Bayesian paradigm. The production
of the equation \eqref{eq:mom} has to occur as a byproduct of defining a joint probability model on the space
$\mathcal{X}\times\Theta$, which seems to logically exclude both non-Bayesian perspectives and {\em ex nihilo}
occurrences of moment conditions. The only statistics example worked out in the paper, namely habit persistence asset
pricing, starts with a given prior distribution (31), which makes this example irrelevant for the stated goal of
checking whether or not ``the assertion of a distribution for moment functions either partially or completely specifies
the prior".

Despite these difficulties in apprehending the paper postulate, I would like to conclude with a more positive
perspective, namely that the problematic of partially defining models and priors via moment conditions is of
considerable interest in an era of Big Data, small worlds \citep{savage:1954}, and limited information. Acknowledging
that inference and in particular Bayesian inference cannot always handle big worlds (see, e.g., the paradox exposed in
\citealp{robins:wasserman:2000}) and constructing coherent and efficient tools for restricted inference about some
aspects of the model are very current questions that beg addressing in full generality.

\section*{Acknowledgements}
I am quite grateful to Robert Kohn (UNSW) and Mark Steel (University of Warwick) for helpful comments and references.
This discussion was first delivered at the 6th French Econometrics Conference, on Dec. 5, 2014, conference held 
in honour of Christian Gouri\'eroux, to whom I am indebted for his help and support throughout my academic career.

\hyphenation{Post-Script Sprin-ger}

\end{document}